\documentclass{article}

\usepackage{PRIMEarxiv}
\usepackage{amsthm}
\usepackage{amstext}
\usepackage{graphicx}
\usepackage{subcaption}
\usepackage{multirow}
\usepackage[square,numbers]{natbib}

\usepackage[utf8]{inputenc} 
\usepackage[T1]{fontenc}    
\usepackage{hyperref}       
\usepackage{url}            
\usepackage{booktabs}       
\usepackage{amsfonts}       
\usepackage{nicefrac}       
\usepackage{microtype}      
\usepackage{lipsum}
\usepackage{fancyhdr}       
\usepackage{graphicx}       
\graphicspath{{media/}}     

\begin{document}
  
\title{Exploring ordered patterns in the adjacency matrix for improving machine learning on complex networks}

\author{
  Mariane B. Neiva, Odemir M. Bruno \\
 *Scientific Computing Group \\ Institute of Physics of Sao Carlos \\ University of Sao Paulo \\
  \texttt{marianeneiva@usp.br, bruno@ifsc.usp.br}}
  
\maketitle

\begin{abstract}
The use of complex networks as a modern approach to understanding the world and its dynamics is well-established in literature. The adjacency matrix, which provides a one-to-one representation of a complex network, can also yield several metrics of the graph. However, it is not always clear that this representation is unique, as the permutation of lines and rows in the matrix can represent the same graph. To address this issue, the proposed methodology employs a sorting algorithm to rearrange the elements of the adjacency matrix of a complex graph in a specific order. The resulting sorted adjacency matrix is then used as input for feature extraction and machine learning algorithms to classify the networks. The results indicate that the proposed methodology outperforms previous literature results on synthetic and real-world data.

 \end{abstract}


\section{Introduction}
\label{sec:introduction}

The emergence of Big Data has sparked an interest in structuring data as closely as possible to reality and evaluating it to extract knowledge. Traditional data analysis often reduces complex phenomena to a simplified object. However, technological advances and the ability to gather, process, and store larger amounts of data allow us to explore information from various viewpoints. Complex networks, systems that can connect elements based on a specific aspect, fill a gap left by classical science. The capability to create a system with elements and relationships shifts the reductionist approach to an integrative one.

In addition, pattern recognition has been a prominent branch of data science in understanding the world through the perspective of technology. If one were to think of a method that could evolve the benefits of artificial intelligence techniques and integrative data analysis, complex networks would be a natural choice. Pattern recognition in complex networks includes a range of algorithms, such as classification and clustering. Although clustering plays a significant role in the field, classification enables us to recognize diseases, species, structures, and cities, among others. This task is crucial nowadays due to the large amount of data generated that would be too time-consuming and costly to analyze manually. Furthermore, complex networks have a significant advantage for pattern recognition in the era of Big Data, as they can be used to model a wide variety of data, from images to biological systems. It has been demonstrated over the years that most real systems exhibit characteristics of small-world and scale-free networks. The former refers to structures in which elements are connected, on average, by short minimum paths, similar to what occurs in social networks where there is a high probability that a person's friend is also a friend of the person in question. The latter refers to the knowledge that there are frequently reached elements in a network, such as prominent researchers in a field or influential articles in a text. The latter example illustrates the importance of using graphs to analyze the patterns and structures of a given organization. Therefore, this work's quantitative analysis focuses on classifying synthetic and real networks.

Based on the advantages mentioned above, researchers have successfully used the model for pattern recognition in various applications, such as the classification of static and dynamic texture \cite{gonccalves2015complex}, shapes \cite{ribas2018distance}, authorship \cite{machicao2018authorship}, and others. Recently, some works such as the use of cellular automata in \cite{miranda2016exploring}, the construction of multidimensional and deep embeddings from networks in \cite{scabini2022deep}, the analysis of angles formed in the graph of shapes in \cite{scabini2017angular,ribas2022complex}. The works have distinguished themselves from traditional analysis that uses classical statistical metrics such as degree and clustering coefficient to create a one-dimensional graph representation. The recent efforts of some researchers to find novel techniques that overcome the redundant information found in the composition of descriptors based on classical metrics have produced good results in graph classification. As shown in \cite{costa2004complex}, the concatenation of some correlated metrics is sometimes not helpful for pattern recognition. However, have we exhausted all the most straightforward analyses on complex networks? This study aims to partially answer this question by investigating a simple alternative to represent the graph: the adjacency matrix.

The adjacency matrix of a graph has a one-to-one correspondence with the graph itself. This representation allows for the quick computation of metrics such as degree and cocitation. However, a visual inspection of the adjacency matrix provides little information, as permutations of the rows or columns do not change the underlying graph. To address this issue, we propose an ordination of the rows of the matrix such that patterns within the matrix become consistent and allow for the distinction of labels in synthetic and real networks. In addition, we evaluate various feature extraction methods applied to the sorted adjacency matrix for classification purposes, including data projection, deep learning feature extraction, CLBP analysis, Hu moments, and classical measurements. Our results on synthetic models, metabolic networks, and social networks demonstrate that our approach can classify networks with over 90\% accuracy, outperforming the accuracy rates of previous works.
        
The paper is structured as follows: Section \ref{sec:proposed} discusses the construction of the adjacency matrix, its significance for complex network analysis, and the proposed ordination. Section \ref{sec:experiment} describes the datasets evaluated in the study and the literature methods used to compare results. In Section \ref{sec:results}, we present two evaluations: first, a visual analysis of synthetic networks to examine whether the proposed ordination highlights important characteristics of the model (see Section \ref{subs:qualitative}). Second, we use various signature methods as descriptors of the networks for classification, and the results are presented in Section \ref{subs:quantitative}. Finally, in Section \ref{sec:conclusion}, we summarize the discussion and highlight the paper's main contributions.

\section{Proposed Approach}\label{sec:proposed}

This section presents an alternative representation to the graph, the adjacency matrix, and describes some complex network metrics.

A complex network is a set G = \{V,E\} where V are the nodes (or vertices) of the system, V = \{$v_1,v_2,v_i,v_j,v_N$\} while E are the edges relating vertices according to an established condition, E = \{$e_{ij} = (v_i, v_j) |$ if $v_i$ and $v_j$ are connected $\}$. As mentioned before, complex networks are derived from graph theory; therefore, the classical definition of the structure is the same as its precursor. However, even though some researchers believe that complex networks are more than just graphs, due to some topological characteristics capable of incorporating complexity into the model, we can (and will) consider both theories the same. Thus, a matrix form represents a graph named adjacency matrix. For an undirected network G with $N$ vertices, the matrix A is computed by:

\begin{equation}
A(v_i,v_j) = \left\{\begin{matrix}
1, \text{ if } v_i \text{ and } v_j \text{ are connected }
\\ 
0, \text{ otherwise} 
\end{matrix}\right.
\end{equation}

It means that every edge $e_{ij}$ and all nodes $v_i$ are represented in A. Furthermore, the adjacency matrix is symmetric for an undirected network, and its diagonal contains only zeros if it has no self-loops. Matrix A can compute a series of statistics such as degree, cocitation ($C = AA^T$), and bibliographic coupling ($B = A^TA$). Cocitation and bibliographic coupling are specially used for directed networks to create a projection and transform the non-symmetric, which are important metrics for the analysis of references and, for instance, to check essential nodes in a network. 

\begin{figure}[h!]
\centering
\caption{A single graph can have several correct representations. While the first image represents the vertices in circles connected by a line, the other two are in the matrix form. As one can check, the visual patterns are different even though both matrices correspond to the same information. Exploiting the distribution of edges in the network would not be reliable. However, a sorting algorithm would allow machine learning methods to classify the patterns.}
\includegraphics[width=0.75\textwidth]{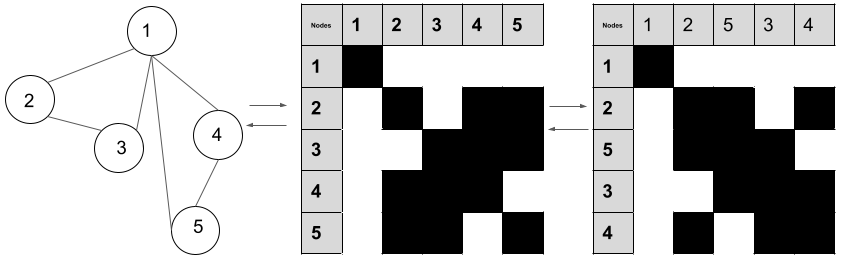}  \label{fig:correspondence}
\end{figure}

However, it can be noticed that in a graph representation, the nodes usually have no order. Therefore,  two adjacency matrices can represent the same network as shown in Figure  \ref{fig:correspondence}. Regarding the lack of spatial order in the matrix, this work proposes a sorting technique of the matrix to find patterns in a graph. Nevertheless, the following section aims to explain this transformation on A and the reason for its choice.

\subsection{Adjacency matrix based signature for complex networks}\label{subs:admSignature}

One of the basic characteristics that can be analyzed on a graph is the degree of a vertice $i$, $k_i$. The metric is related to the amount of connection a vertice has with other nodes in the graph. Also, a straightforward analysis considers the highest' degree node as the most important in the system. Therefore, the first sort of matrix A is to order all nodes decreasingly based on their degrees. It means that the node with a higher degree will populate the first row and column of A. 

However, several nodes may contain the same degree, and the goal of creating a unique network representation will fall apart. Thus, we propose the following rules:

\begin{enumerate}
\item Order the rows according to a descending sort of the degrees of the nodes. 
\begin{enumerate}
\item If two nodes at rows $i$ and $j$ contain the same degree, sort according the highest betweenness.
\end{enumerate}
\end{enumerate}

This simple ordination will ensure that nodes with more connections conquer the first positions in the new A' matrix. Also, due to the second rule, for a single network, matrix A' will be unique for a given input. Since we are analyzing statistics dependent on the matrix arrangement, we need to certify that the matrix image is always the same, regardless of how $A$ was first initialized. 

A', the ordered version of an adjacency matrix A, will be evaluated in this study in two ways: visually and quantitatively. Finally, the final signature is evaluated based on several descriptor methods described in Section \ref{subs:quantitative}. The model's simplicity, in contrast with the results, shows that there is still room to explore basic representations of systems such as the adjacency matrix.

\section{Experiments}\label{sec:experiment}

\subsection*{Complex networks datasets}\label{subs:dataset}

In this study, we will evaluate the use of adjacency-based signatures in four datasets. Therefore, this section describes the two synthetic and two real networks used for the classification task. In the following explanation, $\bar{k}$ is the average degree, and N is the number of nodes.

\begin{itemize}

\item \textbf{Synthetic Dataset}: in complex networks literature, some models are well-known and help us to encapsulate many real phenomena. Therefore, this dataset is composed of four different classes of graphs:

\begin{itemize}

\item Random networks generated by adding edges based in a probability p = $\bar{k}$/n; 

\item Small-world networks which are first created are regular graphs, and then edges are rewired with probability p = 0.1; 

\item Scalefree models with linear and nonlinear preferential attachment; 

\item Geographical networks. 

\end{itemize}

Each model is filled with networks with different average degrees: $\bar{k}$ = \{4, 6, 8, 10, 12, 14, 16\} and different sizes N = \{500, 1000, 1500, 2000\}. For each model combination, average degree, and size, 100 networks are created, totaling 11200 files.

\item \textbf{Scalefree Networks}: among the scalefree models, several researchers proposed techniques to build networks with the power law distribution. This dataset is built with graphs proposed by Barabási \& Albert\cite{barabasi1999emergence} where linear and non linear networks are generated ($\alpha$ = \{0.5, 1.0, 1.5, 2.0\}) and Dorogovtsev \& Mendes \cite{dorogovtsev2013evolution} totaling 500 networks (separated five classes) with size 1000 and $\bar{k}$ = 8.

\item \textbf{Metabolic Dataset}: metabolic systems can be represented as graphs, which is the case of this dataset. With 43 networks of three different species, the metabolic dataset models the system of 6 archaea, 32 bacteria, and five eukaryotes \cite{jeong2000large}. The resample method from Weka platform \cite{hall2009weka} was applied prior to classification to evaluate stratified data. 

\item \textbf{Social Networks}: In the attempt to separate the structure of the social networks of Google+ and Twitter, 65 graphs of each class from SNAP (Stanford Network Analysis Project) platform \cite{leskovec2014stanford} were extracted. Each network was created by taking the social relationships of several users. The goal is to check whether a simple method can distinguish both structures. 

\end{itemize}

\subsection*{Comparison with other methods}\label{subs:otherM}

Two approaches were considered to compare the proposed approach with results in the literature: the LLNA (Life-Like Network Automata) \cite{miranda2016exploring} and the performance of classical structural metrics analysis. A brief description of the methodologies is presented in the following:

\subsubsection*{Life-Life Network Automata}\label{subs:LLNA}

This recent publication uses the advantages of complex networks and cellular automata-based statistics \cite{miranda2016exploring}. First, the method converts the original network to cellular automata. Then, each cell is set as alive or dead, and dynamic evolution is applied. Feature vectors are produced by computing statistics such as entropy and Lempel-Ziv complexity over the automata's evolution. The method also evaluates synthetic and real systems.

\subsubsection*{Classical structural metrics}\label{subs:cMetrics}

The classical analysis of complex networks includes structural metrics such as degree, clustering, and closeness \cite{newman2003mixing, newman2018networks}. To compare our approach with the basics, we compute a set of measurements and evaluate their capacity to distinguish each class alone and combined. The metrics used are. Moreover, as some metrics are dependent on the number of nodes in the network, a histogram was computed and normalized with a size 500.

\begin{itemize}
\item  \textbf{Degree assortativity} ($pp$): this first metric analyses the similarity between neighbors in a graph based on its degrees. The output is a feature vector of the size N (number of nodes).
\item \textbf{Diameter} ($d$): this single measure for the whole network represents the longest distance found among all shortest paths between pairs of nodes in a graph. It requires a high computation cost since the method first computes all the shortest paths before outputting the metric.
\item \textbf{Closeness} (cl): the fourth metric is the vector $cl$ of size N (number of nodes), where for each position $i$, it represents the inverse of the sum of all short distances between $i$ and all other nodes in the graph.
\item \textbf{Eccentricity} (ecc): in contrast with the closeness, the eccentricity of a node is the highest geodesic distance between a node $i$ and any other node in the network.            
\item \textbf{Betweenness} (bet): another measurement based on shortest paths, the betweenness counts how many times a particular node $i$ is the in all shortest paths of a graph. 
\item \textbf{Degree}: the degree is the simplest yet powerful metric to extract from a network. It represents, for a single node, the number of connections it contains. In the experiments, two types of degree analysis are performed: $\vec{k}$ is the vector where for each $k_i$ it returns the degree of node $i$. 
\item \textbf{Clustering}: A node's clustering coefficient measures how close a vertex's neighbors are to a complete graph (all neighbors connected). A vector ($cc_i$) representing the clustering coefficient of each node is used for evaluation. 
\end{itemize}

\section{Results and discussion}\label{sec:results}

\subsection{Visual analysis of the ordered adjacency matrix of synthetic networks}\label{subs:qualitative}

\begin{figure}[h!]
\centering
\includegraphics[width=0.7\textwidth]{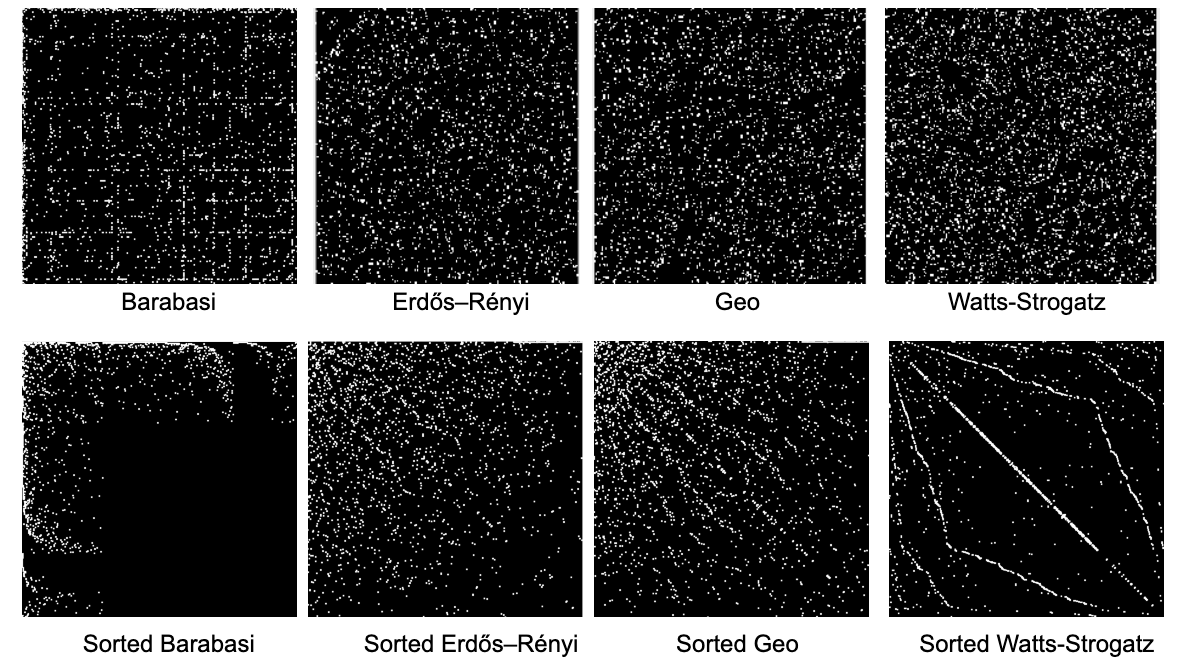}
\caption{The first row shows the unsorted adjacency matrices, while the second shows the result of the proposed approach. The application enhances the features of each model compared to unsorted matrices. White points represent edges, and image pixels are dilated by a small amount to improve visualization.}
\label{fig:example}
\end{figure}

First, we compute the adjacency matrix of all networks of a given dataset and order them to produce matrices where hubs are encountered at the beginning of the graph. The ordination guarantees that the final A' will remain independent of the original arrangement of vertices in the original A. The primary analysis performed in this study is a visual inspection of A'. By analyzing the distribution of white pixels (connections in A'), one can check if the characteristics of synthetic models are enhanced with visual inspection. Figure \ref{fig:example} shows the effect of the methodology in the synthetic dataset (all images are dilated by a small amount to improve visualization).
 
\begin{figure}[htp!]
    \centering
    \begin{subfigure}[b]{0.3\textwidth}
        \includegraphics[width=\textwidth]{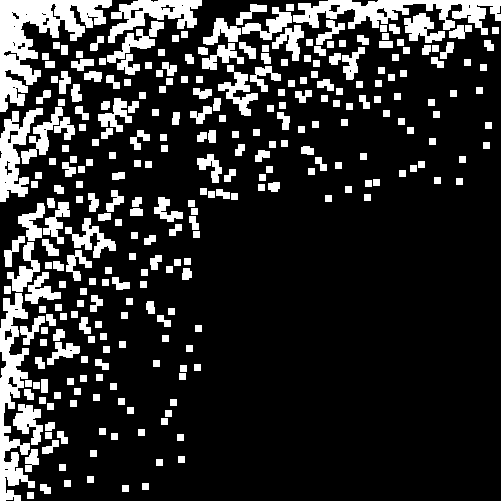}
        \caption{Barabasi, n= 500, $\bar{k}$ =4}
        \label{fig:barabasi_k4}
    \end{subfigure}
    ~ 
    \begin{subfigure}[b]{0.3\textwidth}
        \includegraphics[width=\textwidth]{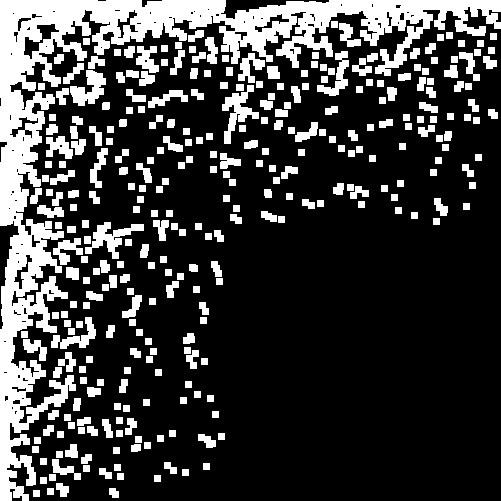}
        \caption{Barabasi, n= 500, $\bar{k}$=6}
        \label{fig:barabasi_k6}
    \end{subfigure}
    ~ 
    \begin{subfigure}[b]{0.3\textwidth}
        \includegraphics[width=\textwidth]{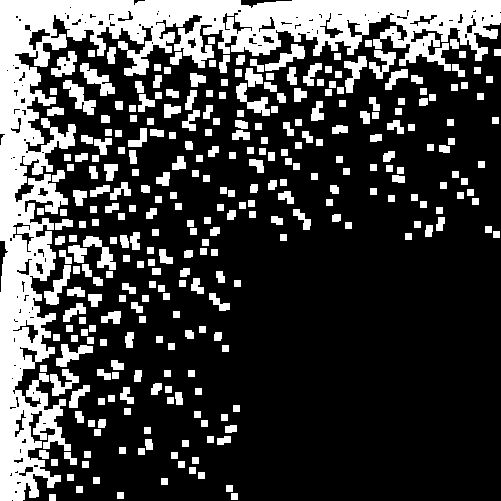}
        \caption{Barabasi, n= 500, $\bar{k}$ =8}
        \label{fig:barabasi_k8}
    \end{subfigure}
    \caption{After the application of the proposed method, the patterns of Barabasi Model become reliable and highlighted}\label{fig:barabasi}
\end{figure}

\begin{figure}[htp!]
    \centering
    \begin{subfigure}[b]{0.3\textwidth}
        \includegraphics[width=\textwidth]{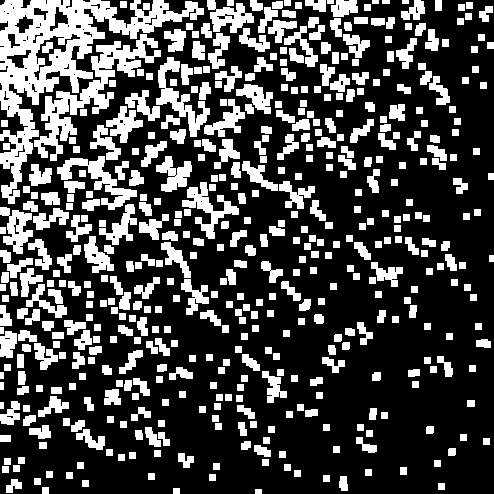}
        \caption{Erdos, n= 500, $\bar{k}$ =4}
        \label{fig:erdos_k4}
    \end{subfigure}
  ~
    \begin{subfigure}[b]{0.3\textwidth}
        \includegraphics[width=\textwidth]{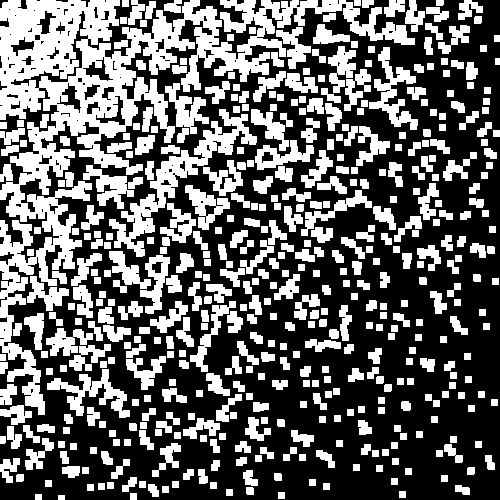}
        \caption{Erdos, n= 500, $\bar{k}$ =6}
        \label{fig:erdos_k6}
    \end{subfigure}
   ~
    \begin{subfigure}[b]{0.3\textwidth}
        \includegraphics[width=\textwidth]{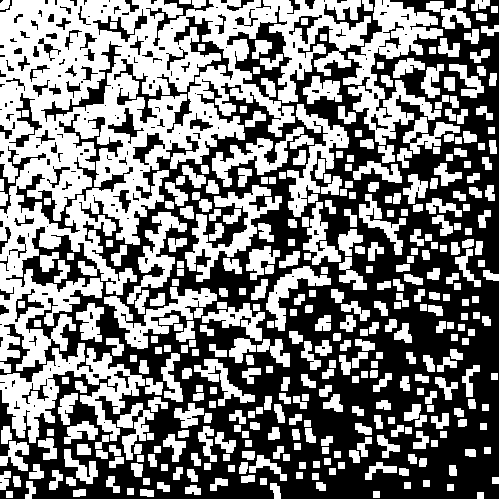}
        \caption{Erdos, n= 500, $\bar{k}$ =8}
        \label{fig:erdos_k8}
    \end{subfigure}
    \caption{The Erdos model images show a wide distribution of points along all the images strengthening the characteristics of the randomness of the network.}\label{fig:erdos}
\end{figure}

\begin{figure}[htp!]
    \centering
    \begin{subfigure}[b]{0.3\textwidth}
        \includegraphics[width=\textwidth]{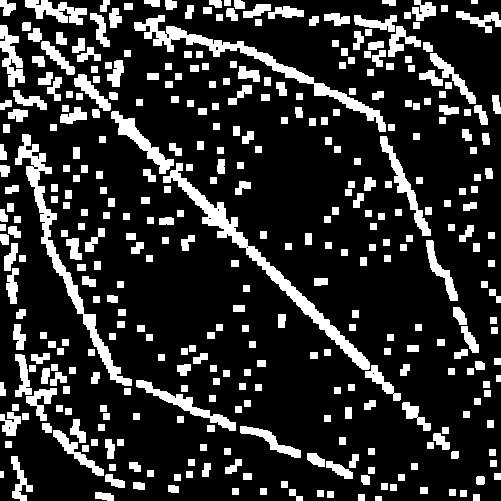}
        \caption{Watts, n= 500, $\bar{k}$ =4}
        \label{fig:watts_k4}
    \end{subfigure}
    ~ 
    \begin{subfigure}[b]{0.3\textwidth}
        \includegraphics[width=\textwidth]{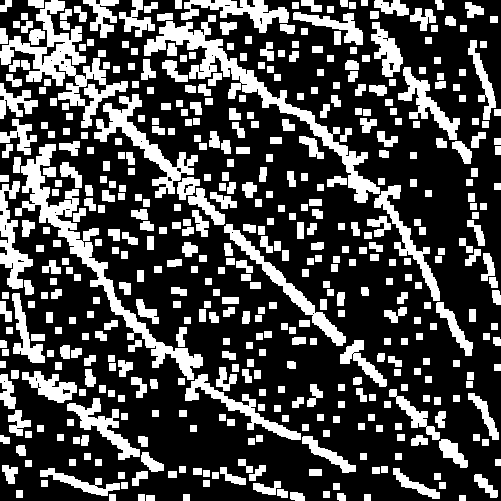}
        \caption{Watts, n= 500, $\bar{k}$ =6}
        \label{fig:watts_k6}
    \end{subfigure}
    ~ 
    \begin{subfigure}[b]{0.3\textwidth}
        \includegraphics[width=\textwidth]{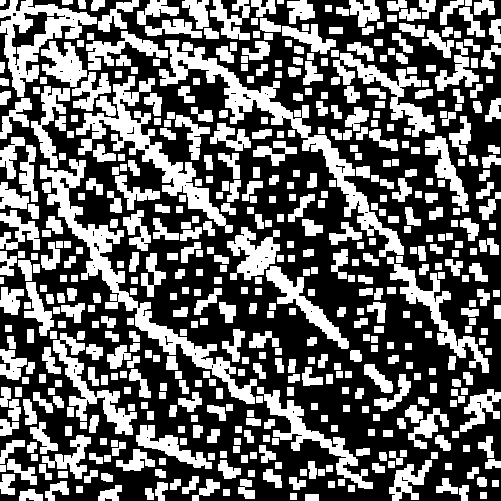}
        \caption{Watts, n= 500, $\bar{k}$ =8}
        \label{fig:watts_k8}
    \end{subfigure}
    \caption{It is interesting to notice how patterns of hubs are represented in Watts model images. Even with the increase in the number of nodes, the basic pattern remains. }\label{fig:watts}
\end{figure}

\begin{figure}[htp!]
    \centering
    \begin{subfigure}[b]{0.3\textwidth}
        \includegraphics[width=\textwidth]{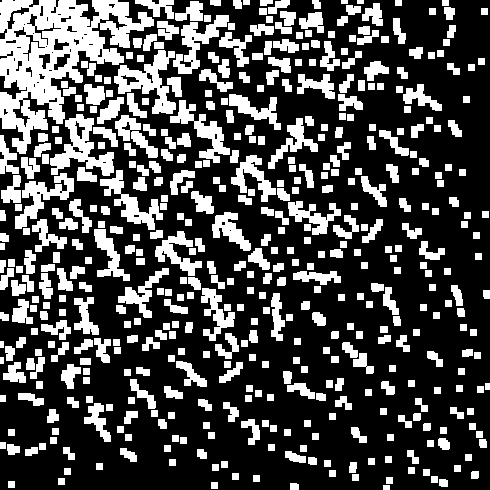}
        \caption{Geo, n= 500, $\bar{k}$ =4}
        \label{fig:geo_k4}
    \end{subfigure}
    ~ 
    \begin{subfigure}[b]{0.3\textwidth}
        \includegraphics[width=\textwidth]{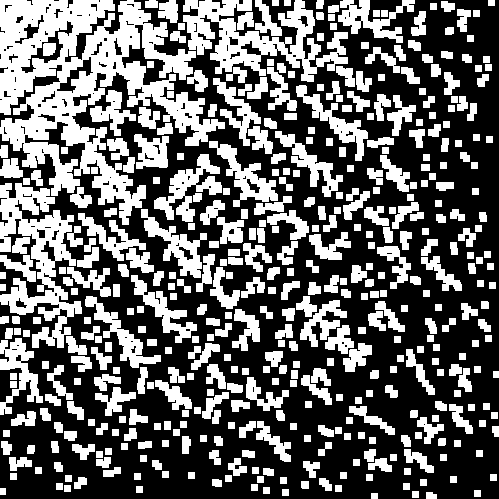}
        \caption{Geo, n= 500, $\bar{k}$ =6}
        \label{fig:geo_k6}
    \end{subfigure}
    ~ 
    \begin{subfigure}[b]{0.3\textwidth}
        \includegraphics[width=\textwidth]{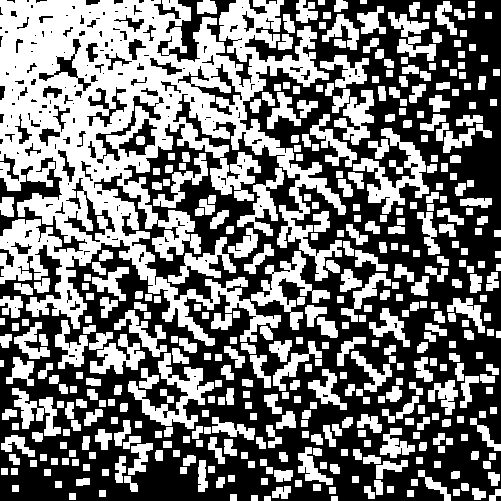}
        \caption{Geo, n= 500, $\bar{k}$ =8}
        \label{fig:geo_k8}
    \end{subfigure}
    \caption{The Geo model is the only one that does not have a solid protocol to be generated. In the images, one could check that, although the lack of strict rules, the patterns within the model maintains.}\label{fig:geo}
\end{figure}

Figures \ref{fig:barabasi}, \ref{fig:erdos}, \ref{fig:watts} and \ref{fig:geo} show some samples from each synthetic class ordered as suggested in Section \ref{sec:proposed}. From visual inspection, we can see some of the theoretical characteristics of these four methods printed in the ordination. 

First, Figure \ref{fig:barabasi} shows examples of Barabasi-Albert networks which is a method to produce scalefree graphs \cite{barabasi1999emergence}. Therefore, to understand the output image, we must understand the model. The scalefree networks proposed by Barabasi-Albert create graphs that initially take an arbitrary network of $N_0$ nodes. Then, nodes are added and linked to $c$, a method parameter, and other vertices. The choice of which vertices are connected to the new node is based on a probability $p = \frac{k_i}{\sum_j k_j}$ where $k_i$ is the degree of vertice $i$. It means that a vertice with a higher degree has a greater chance of being connected with this new node.

Consequently, it is said that there is a preferential attachment between the new vertices and the ones that contain many connections. Consequently, the scalefree networks will present a degree distribution following a power law of the form P(k) ~ $k^{-3}$. Also, an important example of this network is the World Wide Web, where popular websites such as Google are more likely to start a connection with a new page. These heavily connected vertices are known as $hubs$ and are very important to understanding a system and extracting statistics and information about its dynamics.

With this said, we can go back to the visual inspection in Figure \ref{fig:barabasi}. The first aspect to notice is the high density of points in the upper left corner of the images. Each white point represents an edge. Furthermore, the ordination guarantees that nodes with more links are located in the first rows and columns. Also, the lower right is more sparse, showing that the nodes with a lower degree are the majority of the graph, which matches the power law distribution of the model.

Our second model is the Erdos-Renyi network \cite{erdos1960evolution}. The most well-known proposal to create random networks is the following algorithm: $N$ nodes are connected according to a probability $p$. As $p$ is a parameter of the method, graphs with lower $p$ are weakly connected, while a higher $p$ delivers a graph with many edges. The connection between two nodes is performed with a probability independent of the node itself, which is different from what occurs in the previous method. These characteristics are visible in Figure \ref{fig:erdos}; in this case, the accumulation of points only in the upper left side, such as what is in the Barabasi-Albert model, is not found. The points are spread all over the image, giving an idea of uniformity compared to other synthetic models. As expected, the upper left contains more points due to the ordination, but the difference is smaller in random networks compared to scalefree networks.

The most common model found in natural systems is the small world network. A synthetic model to simulate this category was proposed by \citeauthor{watts1998collective} in \citeyear{watts1998collective}, and it initiates the graph as a regular network with $N$ nodes and average degree $k$ where, first, a node is linked to k/2 neighbors on each side. Then, each edge on the right side of a node is rewired with probability $0 \leq \beta \leq  1$ ($\beta$ is a parameter of the method). In the end, the algorithm delivers well-connected nodes with high coefficient clustering and low diameter of the metric graph. In practice, it means a small number of edges linking any two vertices in the system. Finally, let us analyze the output Figures \ref{fig:watts} that represent the ordered adjacency matrices of three model samples. The images show an interesting pattern for the model; it is possible to see a diamond shape that can be explained by the rewiring method applied in the k/2 neighbors of each node and the probability used ($\beta = 0.1$). The pattern shape also allows us to understand the small-world characteristic. If one node cannot reach another straight in one step, connecting will require a small path. This characteristic is visible by the linear line in the pattern and the knowledge that the initial setup of the network is a regular ring.  

Finally, the last synthetic model does not have a solid statistical characteristic, so we cannot assume much about the patterns presented in Figure \ref{fig:geo}. Geographical networks are constructed based on the spatial properties of the system. Important real examples of this model are airport and road traffic systems. Although the lack of strict rules, the patterns within the model maintains. Also, one can notice a pattern similar to the random networks in Figure \ref{fig:erdos}. The similarity may indicate difficulty distinguishing both classes, which statistical methods can evaluate.  

\subsection{Quantitative analysis based on adjacency matrix signature}\label{subs:quantitative}

The visual inspection showed that the approach proposed in Section \ref{subs:admSignature} was feasible for network characterization. However, numerical analysis was also performed using signatures described in this section. In this context, not only do we evaluate synthetic networks but also real and scalefree models. We compare the results obtained by this simple approach with results available in LLNA \cite{miranda2016exploring} and classical structural metrics. Tables \ref{tab:mineknn} to \ref{tab:strc} show the CCR (corrected classification rate) using KNN (k = 1) and SVM. Also, cross-validation, k-fold (k = 10), is used to analyze generalization. It is important to notice that we evaluate several features extracted from images produced by the ordination of the adjacency matrix of each complex network. The local and global methods are analyzed and discussed in this section, from the simplest to more complex levels of exploration of the images. In summary, different experiments are performed: 

\begin{enumerate}
\item \textbf{Projection}:this simple conversion of the 2D image to a 1D representation is performed by summing the values of each column in the \textit{x} axis. Therefore, since the nodes with higher values are located in the first columns, the feature vectors present a descending characteristic. Without any quantitative analysis, one could think that it could be helpful to distinguish random and scalefree graphs due to the difference in the angular coefficient of the vector. Since networks contain different sizes, the projection vector is set to 2500, and when networks are smaller than this size, the remaining values are set to zero. We understand that adding zeros does not influence the results since it means degrees equal to zero. \\

\item \textbf{CLBP}: The Complete Local Binary \cite{guo2010completed} pattern method is an extension of the classical LBP \cite{ojala2002multiresolution}. Unlike its predecessor, which only analyzes the binary difference among neighbors, the CLBP also uses the difference magnitude and signal information to compute the features. The addition of this information has been proven suitable for many applications such as texture recognition \cite{guo2010completed}. In the experiments, the window among a central pixel used to evaluate the neighbors is 3x3. \\

\item \textbf{Hu Moments}:Hu image moments are comprehended as the weighted average of pixels intensities of a particular region or a function of previous moments. The methods have been used to describe areas of the image, area, and centroid. However, some statistics are known to be invariant to scale, translation and rotation. Proposed in \citeauthor{hu1962visual} (\citeyear{hu1962visual}), the Hu moments are a set of seven features constructed based on a function of other moments and encapsulate information such as inertia around the image's centroid and can distinguish mirrored images. The set of seven features is used as input for KNN and SVM analysis.\\

\item \textbf{VGG-19}: very important nowadays due to their ability to classify images, the deep neural networks are a powerful source of features for classification. Therefore, in this experiment, we use the VGG-19 model with pre-trained weights from the Imagenet dataset \cite{simonyan2014very}. The weights output in the last polling layer of the model are used as input for KNN and SVM in the experiments. The methodology for texture analysis in \cite{condori2021analysis} has been promising.

\end{enumerate}

\begin{table}[]
\caption{Classification of the sorted matrices with KNN (k = 1). }\label{tab:mineknn}
\resizebox{\textwidth}{!}{%
\begin{tabular}{|c|cccc|}
\hline
\multirow{2}{*}{}   & \multicolumn{4}{c|}{KNN, k = 1}                             \\ \cline{2-5} 
                    & Projection   & CLBP          & Hu Moments    & VGG-19       \\ \hline
Synthetic           & 96.44 (0.56) & 96.36(0.56)   & 99.68( 0.20)  & 99.29(0.28) \\ \cline{1-1}
Synthetic Scalefree & 99.80 (0.66) & 99.36 (0.94)  & 100.00 (0.00) & 99.80( 0.63)  \\ \cline{1-1}
Social Network      & 84.00 (0.47) & 92.31 (8.88)  & 47.69(15.72)  & 90.00(10.91) \\ \cline{1-1}
Metabolic           & 92.03 (6.09) & 91.95 (11.93) & 97.95(6.20)   & 93.50(10.55) \\ \hline
\end{tabular}
}
\end{table}

\begin{table}[]
\caption{Classification of the ordered matrices by the proposed method with SVM.}\label{tab:minesvm}
\resizebox{\textwidth}{!}{%
\begin{tabular}{|c|cccc|}
\hline
\multirow{2}{*}{}   & \multicolumn{4}{c|}{SVM}                                   \\ \cline{2-5} 
                    & Projection    & CLBP         & Hu Moments   & VGG-19       \\ \hline
Synthetic           & 100.00 (0.00) & 82.00(0.76)  & 53.10( 1.15) & 99.86(0.12)      \\ \cline{1-1}
Synthetic Scalefree & 100.00 (0.00) & 98.74 (1.60) & 88.08 (7.85) & 100.00( 0.00)  \\ \cline{1-1}
Social Network      & 87.69(6.49)   & 92.31 (8.88) & 54.62(11.72) & 91.47( 9.21)  \\ \cline{1-1}
Metabolic           & 91.00(11.74)  & 93.00(11.46) & 80.75(13.84) & 91.50(11.07) \\ \hline
\end{tabular}
}
\end{table}

\begin{table}[]
\caption{Comparison of the proposed method with results in the literature}\label{tab:strc}
\resizebox{\textwidth}{!}{%
\begin{tabular}{|c|cccccccccl|}
\hline
\multirow{2}{*}{} & \multicolumn{10}{c|}{SVM}                                                                                                                                                                      \\ \cline{2-11} 
                  & \multicolumn{1}{c|}{VGG-19}        & \multicolumn{1}{c|}{LLNA}         & $pp$         & $d$          & $cl$         & $ecc$        & $bet$        & $\vec{k}$    & $cc_i$       & combined     \\ \hline
Synthetic         & \multicolumn{1}{c|}{99.86(0.12)}   & \multicolumn{1}{c|}{99.92}        & 77.51(0.47)  & 50.67(1.21)  & 86.95(1.06)  & 73.18(1.43)  & 84.83(1.19)  & 81.97(0.43)  & 92.37(1.07)  & 100.00(0.00) \\ \cline{1-1}
Scalefree         & \multicolumn{1}{c|}{100.00( 0.00)} & \multicolumn{1}{c|}{98.3 (0.3)}   & 93.25(5.28)  & 83.00(3.50)  & 100.00(0.00) & 89.00(5.03)  & 100.00(0.00) & 98.50(2.11   & 100.00(0.00) & 100.00(0.00) \\ \cline{1-1}
Social Network    & \multicolumn{1}{c|}{91.47( 9.21)}  & \multicolumn{1}{c|}{92.00 (1.00)} & 46.92( 7.65) & 44.62( 4.87) & 93.85( 4.87) & 93.85( 4.87) & 86.15( 4.87) & 89.23(6.49)  & 93.85(4.87)  & 92.31( 5.13) \\ \cline{1-1}
Metabolic         & \multicolumn{1}{c|}{91.50(11.07)}  & \multicolumn{1}{c|}{87.00 (13)}   & 75.00(18.26) & 75.00(18.26) & 81.00(17.76) & 83.50(18.57) & 91.00(15.06) & 91.50(14.54) & 84.50(17.23) & 93.50(10.55) \\ \hline
\end{tabular}
}
\end{table}

\subsubsection{Results and Discussion}

Before discussing the results presented in Tables \ref{tab:mineknn} to \ref{tab:strc}, we must remind the proposal's goal. Although the feature extraction methods are not the study's novelty, the subject of analysis is. All methods explained in Section \ref{subs:quantitative} are sorted according to their level of complexity in the analysis of the input. First, the projection, one of the most basic approaches, captures the degree structure of the network since the ordination proposed sorts the nodes according to their degree values. Then, classical metrics, widely used in the literature to represent the node, essentially analyze the graph's structure and how elements connect. Notice that most of the measures used in the experiments rely on the study of shortest paths which can input redundancy in the combined analysis \cite{costa2004complex}. Then, the CLBP, a texture extraction method, analyzes the image locally and can extract important features as the exploration move away from the upper left of the image, where usually a concentration of hubs is encountered. Hu moments, a method invariant to scale, rotation, and translation, analyze statistics of regions in the image, such as inertia and mirroring. It shows important results based on the ability to quantify the global dispersion of the pixels in the proposed adjacency matrix. Finally, VGG-19 is the only method where it is not possible to clearly understand which features are extracted from the patterns in the image. However, deep neural networks build hierarchical features and can create strong representations of the input images, resulting in high classification rates, as seen in the results. 

In the experiments, the two synthetic datasets are constructed based on theoretical constraints, which are well captured by basic statistics. This knowledge is shown by the high classification rate obtained by structural statistics and projection over synthetic and scalefree datasets. It is possible to achieve 100\% in the classification with the combination of structural metrics (see Tables \ref{tab:minesvm} to \ref{tab:strc}). The projection of A' in the $x$ axis can also distinguish very well the models due to the degree distribution. For scalefree synthetic networks, the result of the use of clustering coefficient reached the highest possible result for classification, which shows the ability to distinguish this dataset easily. Finally, for the synthetic networks, the seven Hu moments were also robust to represent the patterns with KNN and output a correct classification rate (CCR) of 100\%. Finally, the deep neural network was also powerful to describe the important features of the model(99\% of accuracy when SVM classifies the features). Therefore, not much more effort has been left knowing that simple metrics easily recognize these sets' structures. 

\begin{figure}[h!]
\centering
\caption{Confusion matrix and area under curve(AUC) of synthetic dataset. The results are obtained by extracting features of VGG-19 neural network and classify then using SVM. }
\includegraphics[width=0.7\textwidth]{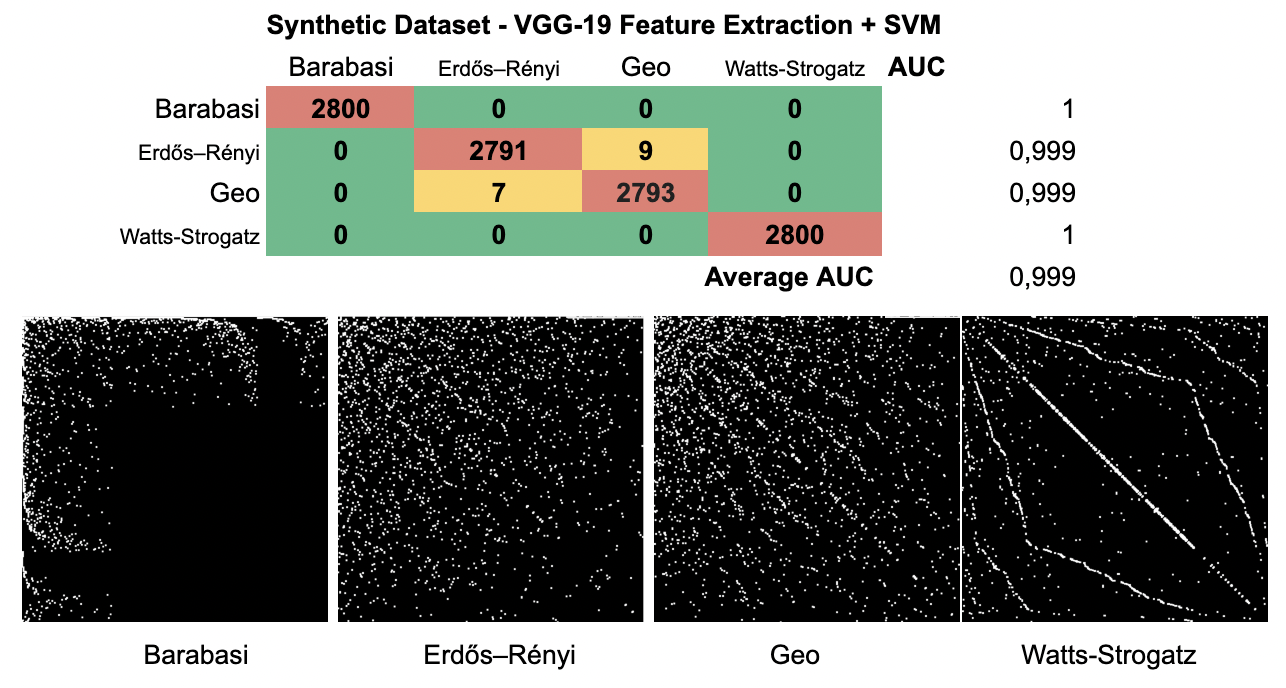}\label{fig:synthetic_auc} 
\end{figure}

\begin{figure}[h!]
\centering
\caption{Confusion matrix and AUC of scalefree dataset for VGG-19 and SVM classification.}
\includegraphics[width=0.9\textwidth]{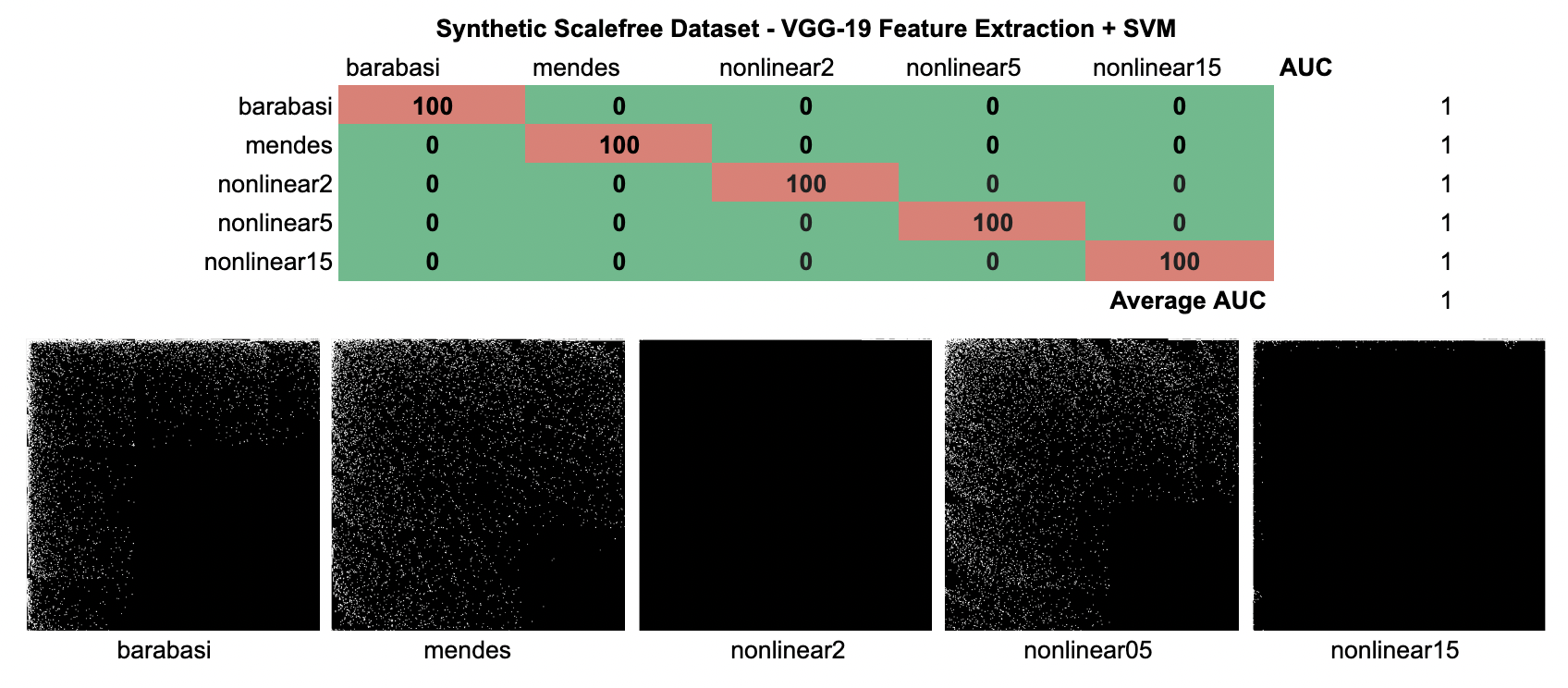}\label{fig:scalefree_auc} 
\end{figure}

Concerning the confusion matrices of VGG-19 feature extraction and SVM classification of the proposal in Figures \ref{fig:synthetic_auc} and\ref{fig:scalefree_auc}. The results navigate into the urge of understanding the accuracy of 99.86\% and 100\% of the synthetic and scalefree datasets, respectively, presented in Table \ref{tab:strc}. As expected from the visual inspection results, the Geo and Erdos-Renyi are mistaken due to their less distinct structural features. Erdos-Renyi and geographic networks had been confused 16 times: seven geographic networks were predicted as Erdos-Renyi, and nine geographic were classified as random graphs. On the other hand, Barabasi and Watts-Strogatz maintain the maximum accuracy in the classification. For the scalefree dataset, all highlighted patterns are beneficial for the classification.

However, knowing that the approaches evaluated work for theoretical analysis, it is important to explore the real world. Two datasets of different natures are studied: metabolic and social networks. It is possible to notice that even the simple projection returns good classification rates, emphasizing the importance of the proposed process for adjacency matrix exploitation. Hu Moments' seven features distinguished the different metabolic networks with a correct rate of 97.95\% (using KNN in Table \ref{tab:mineknn}). Furthermore, the Complete LBP recognizes both networks with over 90\% of CCR, and even basic structural metrics (combined or not) can reach similar recognition rates. 

However, the limitation with simple hand-craft descriptors such as projection and local pattern extractors is that the standard deviation is very high (over 6\% in the experiments). It occurs for the proposed approaches but also for structural and LLNA methods. Also, one disadvantage of LLNA is the parameter setup. The method works by transforming the original network into a cellular automata. Then, the approach evolves the tessellation based on a Life-life rule. However, finding the best rule to achieve good classification takes much work and requires a high computational cost. It means that, although there is a promise of achieving good results with this approach, the disadvantages of computation cost and parameter setup puts the simple approach of ordering the adjacency matrix in favor.

\begin{figure}[h!]
\caption{Confusion matrix and AUC of social dataset for VGG-19 and SVM classification.}
\centering
\includegraphics[width=.7\textwidth]{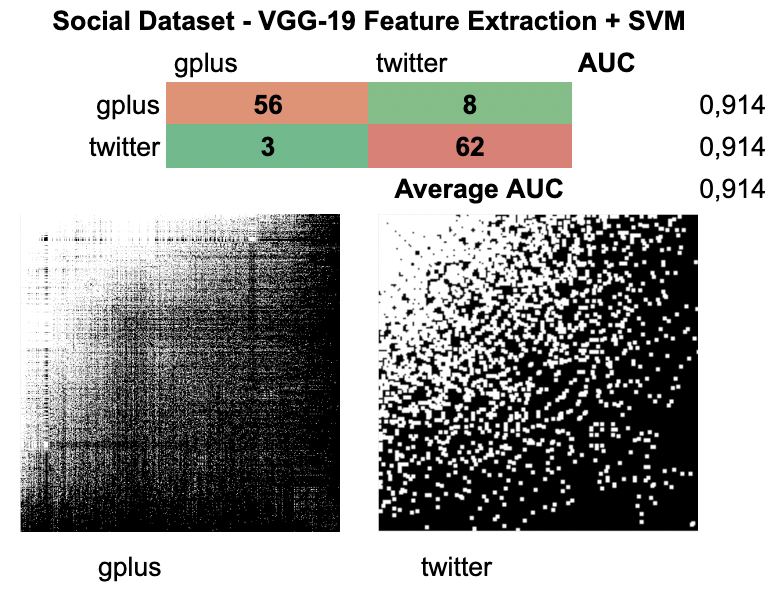}\label{fig:social_AUC}
\end{figure}

\begin{figure}[h!]
\centering
\caption{Confusion matrix and AUC of metabolic dataset for VGG-19 and SVM classification.}
\includegraphics[width=0.7\textwidth]{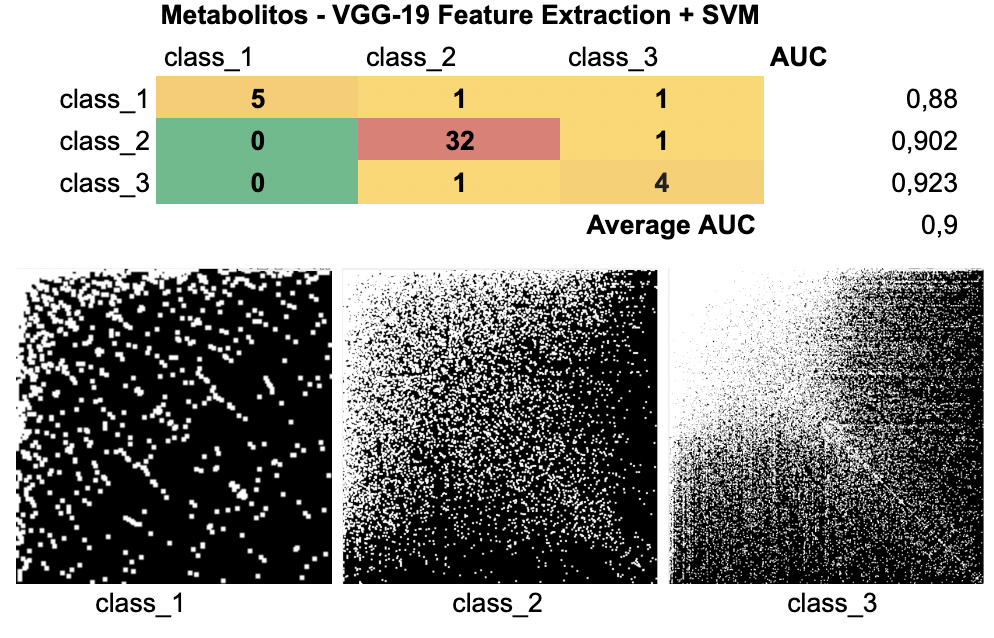}\label{fig:metabolitos_AUC}
\end{figure}

The VGG-19 combined with SVM could classify metabolic networks with 91.50\% and 91.47\% of CCR for social graphs. In addition, confusion matrices show the exact mistakes of classification. Gplus and Twitter had been confused 11 times: eight Gplus networks were predicted as Twitter, and three were misclassified on the opposite side. 

For the metabolic dataset, the class\_2 and class\_3 output one mistake for each class. For class\_1, two images were incorrectly labeled. On average, the AUC is over 0.9 for each evaluated set, showing the methodology's importance.
  
\section{Conclusion}\label{sec:conclusion}

This paper presents the study of exploring a simple feature of the graph as the main input of analysis. The adjacency matrix has a one-to-one representation of the network, but a simple permutation of the rows will keep all system properties non-ordered. Therefore, the paper proposes an ordination of the matrix based on the degrees of the nodes. Therefore, a network can generate a single representation that can be analyzed as an image. Visual analysis was performed, showing that the proposed transformation of the data emphasizes the patterns within the models. Also, quantitative analysis is evaluated in four different datasets, including two real systems, and results show that, in general, all proposed approaches and compared methods output good classification accuracies.   

However, there are three aspects to analyze: correct classification rate, standard deviation, and complexity of the method. It is clear that for synthetic data, basic metrics are enough to represent the networks. Nevertheless, for real datasets, a high standard deviation is usually obtained due to the variability of the data. LLNA has as disadvantages the long time necessary to fit the parameters. Therefore, the power of deep neural networks for image classification (the adjacency matrix can be understood as an image in this experiment) arises again and show promising results mainly in the two later sets of metabolic and social networks. In conclusion, the proposed approach of sorting and analyzing a basic network representation is suitable for further studies with practical domains. It shows us that a simple representation, such as the adjacency matrix, is a good source of information for network classification.

\section*{Acknowledgements}
O. M. Bruno acknowledges support from CNPq (Grant \#307897/2018-4) and FAPESP (grants \#2018/22214-6 and \#2021/08325-2). M. B. Neiva recognizes the support from CNPq (Grant PROEX-9056169/D). The authors are also grateful to the NVIDIA GPU Grant Program.


\end{document}